\documentclass[final]{aipproc}
\usepackage{makeidx}
\usepackage{epsfig}
\usepackage{graphicx}

\input{aipcheck}
\layoutstyle{8x11single}
\graphicspath{{./fig1/}}

\begin{document}

\title{Relativistic kinetic theory of magnetoplasmas}
\author{Alexei Beklemishev\thanks{email:
beklemishev@inp.nsk.su}$\ \ ^{a}\ $, Piero Nicolini\thanks{email:
Piero.Nicolini@cmfd.univ.trieste.it}$\ \ ^{b,c}\ $ \ and Massimo
Tessarotto\thanks{email: M.Tessarotto@cmfd.univ.trieste.it}$\ \
^{b,d} $} {address={\ $^{a} $Budker Institute of Nuclear Physics,
Novosibirsk, Russia\\
\ $^{b} $Department of Mathematics and Informatics,
University of Trieste, Italy\\
$^{c} $National Institute of Nuclear Physics (INFN), Trieste
Section, Italy\\
$^{d} $Consortium for Magnetofluid Dynamics\thanks{Web site:
http://cmfd.univ.trieste.it}, University of Trieste, Italy}}

\begin{abstract}
Recently, an increasing interest in astrophysical as well as
laboratory plasmas has been manifested in reference to the
existence of relativistic flows, related in turn to the production
of intense electric fields in magnetized systems \cite{Mohanti}.
Such phenomena require their description in the framework of a
consistent relativistic kinetic theory, rather than on
relativistic MHD equations, subject to specific closure
conditions. The purpose of this work is to apply the relativistic
single-particle guiding-center theory developed by Beklemishev and
Tessarotto \cite{Beklem}, including the nonlinear treatment of
small-wavelength EM perturbations which may naturally arise in
such systems \cite{Beklem2004}. As a result, a closed set of
relativistic gyrokinetic equations, consisting of the
collisionless relativistic kinetic equation, expressed in hybrid
gyrokinetic variables, and the averaged Maxwell's equations, is
derived for an arbitrary four-dimensional coordinate system.
\end{abstract}

\maketitle



\section{Introduction}
A basic prerequisite for the formulation of a consistent
relativistic kinetic for strongly magnetized plasmas in
astrophysical problems, is the formulation of single-particle
dynamics in the context of a relativistic, fully covariant,
formulation of gyrokinetic theory
\cite{Beklem,Beklem2004,Pozzo1998}. As is well known, this regards
the so-called ``gyrokinetic problem'', i.e., the description of
the dynamics of a charged particle in the presence of suitably
``intense'' electromagnetic (EM) fields realized by means of
appropriate perturbative expansions for its equations of motion.
The expansions, usually performed with respect to the ratio
$\varepsilon =r_{L}/L<<1$, where $L$ and $r_{L}$ are respectively
a characteristic scale length of the EM fields and the
velocity-dependent particle Larmor radius $r_{L}=\frac{w}{\Omega
_{s}},$ with $\Omega _{s}=\frac{qB}{mc}$ the Larmor frequency and
${\bf w}$ the orthogonal component of a suitable particle
velocity. The goal of gyrokinetic theory is to construct with
prescribed accuracy in $\varepsilon $ the so called
``gyrokinetic'' or ``guiding center variables'', by means of an
appropriate ``gyrokinetic'' transformation, such that the
equations of motion result independent of the gyrophase $\phi $,
$\phi $ being the angle of fast gyration, which characterizes the
motion of a charged particle subject to the presence of a strong
magnetic field. In non-relativistic theory the gyrokinetic
transformation can be constructed by means of a perturbative
expansion of the form:
\begin{equation}
{\bf z}{\bf \rightarrow z}^{\prime }={\bf z}_{0}^{\prime }+\varepsilon {\bf z%
}_{1}^{\prime }+\varepsilon ^{2}{\bf z}_{2}^{\prime }+..,
\end{equation}
which in terms of the Newtonian state ${\bf x=(r,v)}$ reads:
\begin{equation}
{\bf r}={\bf r}^{\prime }+\varepsilon {\bf \rho }^{\prime }+\varepsilon ^{2}%
{\bf r}_{2}^{\prime }({\bf z}^{\prime },t,\varepsilon ),
\end{equation}
\begin{equation}
{\bf v}=u^{\prime }{\bf b}^{\prime }+{\bf w}^{\prime }+{\bf
V}^{\prime }+\varepsilon {\bf v}_{1}^{\prime }({\bf z}^{\prime
},t,\varepsilon ),
\end{equation}
where $\varepsilon {\bf \rho }^{\prime }$ is the Larmor radius,
\begin{equation}
\varepsilon {\mathbf \rho }^{\prime }=-\varepsilon
\frac{{\bf w}^{\prime }\times {\bf b}^{\prime }}{\Omega
_{s}^{\prime }},
\end{equation}
while ${\bf V}^{\prime }$ is the  electric drift velocity
\begin{equation}
{\bf V}^{\prime }=\frac{c{\bf E}^{\prime }\times {\bf B}^{\prime }}{%
B^{\prime 2}},
\end{equation}
and the phyrophase gyrophase $\phi ^{\prime }$ is defined:
\begin{equation}
\phi ^{\prime }=arctg\left\{ \frac{({\bf v}^{\prime }-{\bf
V}_{s}^{\prime
})\cdot \widehat{{\bf e}}_{2}^{\prime }}{({\bf v}^{\prime }-{\bf V}%
_{s}^{\prime })\cdot \widehat{{\bf e}}_{1}^{\prime }}\right\} .
\end{equation}
In the past several methods have been devised to construct hybrid
gyrokinetic variables. These include perturbative theories based,
respectively, on non-canonical methods (see for example
\cite{Morozov 1966}), canonical perturbation theory
\cite{Gardner1959,Weitzner 1995}, canonical and non-canonical
Lie-transform approaches \cite{Littlejohn1979,Littlejohn1981}, as
well as Lagrangian non-canonical formulations which make use of
the so-called {\it hybrid Hamilton variational principle}
\cite{Littlejohn1983,Pozzo1998,Beklem}.

  \begin{figure}[!htbp]
      \leavevmode
      \includegraphics[height=2.0in]{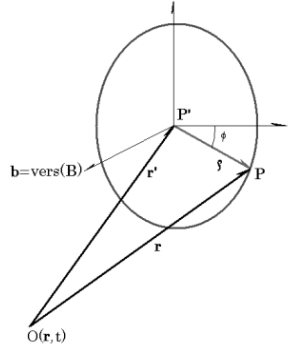}
      \caption{Guiding center and Larmor radius in
non-relativistic theory.  Here $\lbrack \widehat{{\bf e}}_{1}^{\prime },\widehat{{\bf e}}_{2}^{\prime },%
\widehat{{\bf e}}_{3}^{\prime }={\bf b}^{\prime }\rbrack$ denotes
a right-handed basis of unit vectors.}

  \end{figure}

\section{Relativistic gyrokinetic theory: motivations}
\noindent For a certain class of problems in plasma-physics and
astrophysics, existing limitations of the standard gyrokinetic
theory \citep{LJ,BO,CO,Br} make its use difficult or impossible.
In particular, this involves the description of experiments in
which the electric field may become comparable in strength to the
magnetic field (so that the drift velocity becomes relativistic),
and the study of relativistic plasma flows in gravitational
fields, which are observed or assumed to exist in accretion disks
and related plasma jets around neutron stars, black holes, and
active galactic nuclei. The finite Larmor radius effects and the
influence of short wavelength electromagnetic perturbations are
also expected to play a fundamental role in particle dynamics. In
many respects, previous relativistic theory results inadeguate for
such a task. In fact, some of mearlier treatments consider the
particle velocity as relativistic while its drift velocity is not
\cite{Grebogi,Littlejohn1985,Boozer1996,Cooper1997,Brizard1999}.
 This deficiency has been pointed out by Pozzo and Tessarotto
\cite{Pozzo1998}, who developed a special-relativistic theory
which includes the effect of relativistic drifts. However, the
self-consistent inclusion of the gravitational fields, a
prerequisite to make the theory suitable for astrophysical
applications, as well the treatment of nonlinear EM perturbations
of relativistic plasmas, requires a relativistic covariant
formulation. This has been investigated by Beklemishev and
Tessarotto \cite{Beklem,Beklem2004}. In this theory the
appropriate relativistic gyrokinetic theory has been carried out
through second order in the expansion parameter, including wave
fields, based on a Lagrangian approach making use of the hybrid
Hamilton variational principle. In such a case the variational
functional for a charged point particle with the rest-mass $m_{a}$
and charge $q_{a}$ in prescribed fields can be written:
\begin{equation}
S{=}\int {Q_{\mu }\mathrm{d}x^{\mu }=}\int (qA_{\mu }(x^{\nu })+u_{\mu })%
\mathrm{d}{x^{\mu }},  \label{S1}
\end{equation}%
where $q=q_{a}/m_{a}c^{2},$ and variations of $u_{\mu }$ occur on
the seven-dimensional hypersurface $u_{\mu }u^{\mu }=1$, being
$u_{\mu }$ the four-velocity $u^{\mu }=\frac{dx^{\mu }}{ds}$ and
the relevant tensor notations are standard. Thus, $g_{\mu \nu }$
denotes the metric tensor components, characterizing the
coordinate system (and the underlying space-time structure) which
provides the connection between the co- and countervariant
components of four-vectors (henceforth referred to as 4-vectors)
$A_{\mu }=g_{\mu \nu }A^{\nu }$, while the invariant interval $%
\mathrm{d}s$ is defined as
\begin{equation}
\mathrm{d}s^{2}=g_{\mu \nu }\mathrm{d}x^{\mu }\mathrm{d}x^{\nu },
\label{ds}
\end{equation}%
where the Greek indices are assumed to go through $\mu ,\nu
=0,...,3$.

\section{The relativistic gyrokinetic transformation}
The object of gyrokinetics is to introduce a new set of
phase-space variables (called the ``gyrokinetic variables'') such
that the variable
describing the rotation angle along the Larmor orbit (i.e. the gyrophase $%
\phi $) becomes ignorable. This happens, by definition, when the
Lagrangian (or, more generally, the functional) is independent of
$\phi $. Once an ignorable variable is found, the number of
corresponding Euler equations is reduced by one, and the new
variables allow simplified numerical calculations, as the motion
is effectively integrated over the fast Larmor rotation. The
one-to-one transformation from the original set of phase-space
variables $(x^{\mu },u^{\nu })$ to the gyrokinetic variables is
called the ``gyrokinetic transformation''. In what follows, we use
the Lagrangian perturbative approach to find those variables,
which is equivalent (in broad terms), to the Lie-transform method,
though more direct.

First, we assume that the curvature radius of the space-time and
the gradient lengths of the background electromagnetic fields are
much larger than the Larmor radius characterizing the particle
path. However, we allow for existence of wave-fields with sharp
gradients [$k\rho _{L}\sim O(1),$
including $k_{\Vert }\rho _{L}\sim O(1)$,] and rapidly varying in time [$%
\omega \rho _{L}/c\sim O(1)$], while such fields are assumed
suitably smaller in strength than the background field. (We stress
that, unlike in conventional formulations of gyrokinetic theory,
this type of ordering is required in a covariant theory due to the
reference-frame dependence of the ordering assumptions involving
space and time scale lengths of the perturbations.) For this
purpose we introduce the ordering scheme following the notation of
\cite{LJ}:
\begin{equation}
{Q_{\mu }\mathrm{d}x^{\mu }}=\{u_{\mu }+q(\frac{1}{\varepsilon
}A_{\mu }+\lambda a_{\mu })\}\mathrm{d}{x^{\mu }},  \label{L2}
\end{equation}%
where $\varepsilon $ and $\lambda $ are formal small parameters
(they should be set to 1 in the final results) allowing
distinction between the large-scale background field $A_{\mu },$
and the wave-fields given by $ a_{\mu }$. We search for the
gyrokinetic transformation $(y^{i})\equiv (x^{\prime \alpha },\phi
,\widehat{\mu },{u_{\parallel }})\leftrightarrow
(x^{\alpha },u^{\beta })$ in the form of an expansion in powers of $%
\varepsilon $:
\begin{equation}
x^{\nu }=x^{\prime \nu }+\sum_{s=1}\varepsilon ^{s}r_{s}^{\nu
}(y^{i}), \label{x'}
\end{equation}%
where $\phi $ is the ignorable phase variable (gyrophase),
$\widehat{\mu }$ and ${u_{\parallel }}$ represent two other
independent characteristics of velocity (to be defined later),
$x^{\prime \nu }$ is the 4-vector ``guiding center'' position,
$\mathbf{{r}_{s}}$ are arbitrary 4-vector functions of
the new variables $(y^{i})$ to be determined. We require that $\mathbf{{r}%
_{s}}$ are purely oscillatory in $\phi $, i.e., the $\phi $-averages of $%
\mathbf{{r}_{s}}$ are zero, as a part of the $x^{\prime \nu }$-
definition. Note that the above descriptions of the new variables
$(x^{\prime \alpha },\phi ,\widehat{\mu },{u_{\parallel }})\ $will
acquire precise mathematical meaning only as a result of the
search for the gyrokinetic transformation.

This search consists in applying the expansion (\ref{x'}) to the
fundamental 1-form (\ref{L2}) and imposing the requirement that it
is independent of $\phi $ in each order. A convenient framework is
provided by projecting all 4-vectors and 4-tensors along the
directions of a suitable fundamental tetrad $(\tau ,l,l^{\prime
},l^{\prime \prime }) $: i.e., an orthogonal basis of unit
4-vectors so that the last three are space-like, and
\begin{equation}
\sqrt{-g}e_{\varsigma \lambda \mu \nu }\tau ^{\varsigma
}l^{\lambda }l^{\prime \mu }l^{\prime \prime \nu }=1,
\label{orderb}
\end{equation}%
where $\sqrt{-g}e_{\varsigma \lambda \mu \nu }$\ is the purely
antisymmetric tensor. As a consequence the four-velocity can be
represented in the form:
\begin{equation}
u_{\mu }=w\left( l_{\mu }^{\prime }\cos \phi +l_{\mu }^{\prime
\prime }\sin \phi \right) +\bar{u}_{\mu },  \label{um}
\end{equation}%
which can be also regarded as the definition for the gyrophase $\phi $:%
\textit{\ }it is defined as\textit{\ an angle in the velocity-subspace, }%
where we introduce the cylindrical coordinate system. This
definition is covariant with respect to transformations of the
space-time coordinate system, which may change the vector
components, but not the vectors themselves. Furthermore, we assume
that $w$ and $\bar{u}_{\mu }$ are independent of $\phi .$ Validity
of this assumption is justified by existence of the solution (at
least for a non-degenerate Faraday tensor).

The $\phi $-independent part of the 4-velocity
$\mathbf{{\bar{u}}}$ is not completely arbitrary, but satisfies
certain restrictions following from the requirement $u_{\mu
}u^{\mu }=1$ for all $\phi $:
\begin{equation}
\bar{u}_{\mu }={u_{\parallel }}l_{\mu }+u_{o}\tau _{\mu },
\label{bu}
\end{equation}%
\begin{equation}
u_{o}^{2}=1+w^{2}+{u_{\parallel }}^{2}.  \label{uo}
\end{equation}%
Any two of three scalar functions $w,u_{o}$ or ${u_{\parallel }}$
can be considered independent characteristics of velocity, while
the third can be expressed via (\ref{uo}). It is now
straightforward to eliminate from ${\delta G}^{\prime }$ terms
oscillating in $\phi $ by properly defining displacements
$\mathbf{r}_{s}.$ This task can, in principle, be carried out
systematically at any order in the relevant expansion parameters
(in particular in $\varepsilon$). Thus, to leading order in
$\varepsilon$ to eliminate the gyrophase-dependent terms in the
fundamental differential 1-form one must impose constraint:
\begin{equation}
\tilde{u}_{\mu }+\lambda q\widetilde{a}_{\mu }-qr_{1}^{\nu }F_{\mu
\nu } =0,  \label{cond}
\end{equation}%
where $\tilde{y}$ denotes the oscillating part of $y$, namely $\tilde{y}=y-%
\bar{y}$, $\bar{y}=\langle y\rangle _{\phi }$ is the
gyrophase-averaged part of $y$ and $F_{\mu \nu }$ is the EM field
tensor. If the above requirement (\ref{cond})is satisfied, the
gyrophase $ \phi $ is ignorable and the hybrid variational
principle in our approximation can be expressed as $\delta
S^{\prime \prime }=0$. As a result, the $\phi $-independent
functional $S^{\prime \prime }$ becomes
\begin{equation}
S^{\prime \prime }=\int \left\{ \left( \frac{q}{\varepsilon
}A_{\mu }^{\prime }+\lambda q\overline{a}_{\mu }+u{_{\parallel
}}l_{\mu }+u_{o}\tau _{\mu }\right) \mathrm{d}x^{\prime \mu
}+\widehat{\mu }\mathrm{d}\phi \right\} \emph{,}  \label{AS}
\end{equation}
where $\hat{\mu}$ is the relativistic wave-field-modified magnetic
moment, accurate to order $\varepsilon ^{1}$ and
$u_{o}=\sqrt{1+w^{2}+u{_{\parallel }}^{2}}.$\\
The equations of motion, expressed in terms of the relationships
between differentials tangent to the particle orbit, can be
obtained as Euler equations of the transformed variational
principle \cite{Beklem,Beklem2004}. Using the $\phi$-independent
functional (\ref{AS}) in the variational principle $\delta
S^{\prime\prime}=0$ defines the particle trajectory in
terms of the new gyrokinetic variables $(x^{\prime \mu}, \hat{\mu}, {%
u_\parallel}, \phi)$. This set is non-canonical, but further
transformations of variables (not involving $\phi$) also lead to
$\phi$-independent functionals and can be used for this purpose.

\section{The relativistic gyrokinetic Vlasov kinetic equation}
The single-particle distribution function can be written in
general relativity either in the eight-dimensional phase space
$\Phi (x^{\mu },u_{\nu }),$ $\ \mu ,\nu =0,...,3,$\ or in the
seven-dimensional phase
space $f(x^{\mu },u_{\nu }),$ where only 3 components of the 4-velocity $%
u_{\nu }$ are independent, so that
\begin{equation}
\Phi (x^{\mu },u_{\nu })=f(x^{\mu },u_{\nu })\delta
(\sqrt{u_{\zeta }u^{\zeta }}-1)\theta (u^{0}).
\end{equation}
The $\delta -$function here reflects the fact that $u_{\zeta
}u^{\zeta }=1$ is the first integral of motion in the case of the
eight-dimensional representation. \\
The kinetic equation in both cases retains the same form and
yields the collisionless Vlasov kinetic equation, namely
\begin{equation}
u^{\mu }\frac{\partial f}{\partial x^{\mu }}+\left( \frac{du_{\nu }}{ds}%
\right) \frac{\partial f}{\partial u_{\nu }}=0,
\end{equation}
although in the 7-dimensional case $\nu =1,2,3$ only, while
$u^{0}$\ is the dependent variable. Here $\left( du_{\nu
}/ds\right) $ is a function of independent variables $(x^{\mu
},u_{\nu })$ found as the right-hand side of the single-particle
dynamics equations. The kinetic equation can be multiplied by
$\mathrm{d}s$\. In this way it can also be represented in the
parametrization-independent form as follows:
\begin{equation}
\frac{\partial f}{\partial x^{\mu }}\mathrm{d}x^{\mu }+\frac{\partial f}{%
\partial u_{\nu }}\mathrm{d}u_{\nu }=0,
\end{equation}
where the differentials are tangent to the particle orbit.\\
Due to general properties of variable transformations it is
obvious that any non-degenerate transformation of the phase-space
variables $(x^{\mu },u_{\nu })\rightarrow (y^{i})$ will lead to
the same form of the kinetic equation
\begin{equation}
\frac{\partial f}{\partial y^{i}}\mathrm{d}y^{i}=0,
\end{equation}%
where the differentials are tangent to the particle orbit. In
particular, this property is useful for transformation to the
gyrokinetic variables.

Let $(y^{i})\equiv (x^{\prime \alpha },\phi ,\widehat{\mu },{u_{\parallel }}%
) $, then the kinetic equation becomes
\begin{equation}
\frac{\partial f}{\partial x^{\prime \mu }}\mathrm{d}x^{\prime \mu }+\frac{%
\partial f}{\partial u_{\parallel }}\mathrm{d}u_{\parallel }+\frac{\partial f%
}{\partial \widehat{\mu }}\mathrm{d}\widehat{\mu }+\frac{\partial f}{%
\partial \phi }\mathrm{d}\phi =0.
\end{equation}%
By definition of the gyrokinetic variables the dynamic equations
should be independent of $\phi $\ , i.e., expressions for \
$\left( dx^{\prime \mu
}/d\phi \right) ,\left( du_{\parallel }/d\phi \right) ,\left( d\widehat{\mu }%
/d\phi \right) $ are independent of $\phi $, while $\partial
f/\partial \phi $ is periodic in $\phi $. It follows that
$\partial f/\partial \phi =0,$ and, if $\widehat{\mu }$ is the
integral of motion, $\mathrm{d}\hat{\mu}=0$, we get the kinetic
equation expressed in the gyrokinetic variables as
\begin{equation}
\frac{\partial f}{\partial x^{\prime \mu }}\mathrm{d}x^{\prime \mu }+\frac{%
\partial f}{\partial u_{\parallel }}\mathrm{d}u_{\parallel }=0,
\label{gyrok}
\end{equation}%
which we shall call {\it relativistic gyrokinetic Vlasov kinetic equation}. Here the coefficients $%
\mathrm{d}x^{\prime \mu }$ and $\mathrm{d}u_{\parallel }$ must be
determined from the equations of motion in the gyrokinetic
variables.

\section{The Maxwell's equations}
Finally we point out another important feature of the present
formulation of the gyrokinetic theory. Namely, the Jacobian of the
transformation is simple enough to allow explicit integration in
the gyrophase, needed for evaluation of the charge and current
densities. The general form of the Maxwell's equations in presence
of an arbitrary gravitational field is well known\cite{LL1}. The
first pair of equations can be written as
\begin{equation}
e^{\varsigma \lambda \mu \nu }\frac{\partial F_{\mu \nu
}}{\partial x^{\lambda }}=0,
\end{equation}%
while the second as
\begin{equation}
\frac{1}{\sqrt{-g}}\frac{\partial }{\partial x^{\nu }}\left(
\sqrt{-g}F^{\mu \nu }\right) =-\frac{4\pi }{c}j^{\mu }, \label{sp}
\end{equation}%
where
\begin{equation}
j^{\mu }=c\sum_{\alpha }q_{\alpha }\int u^{\mu }f_{\alpha }\left( \mathbf{x},%
\mathbf{u}\right) \delta \left( \sqrt{u^{\nu }u_{\nu }}-1\right) \frac{%
\mathrm{d}^{4}\mathbf{u}}{\sqrt{-g}}
\end{equation}%
is the current density, expressed via the distribution function of
particle species $\alpha ,$ and the signed particle charge
$q_{\alpha }$. The $\delta $-function under the integral allows to
make partial integration, for example over $u_{0},$\ and arrive at
a more widely used form
\begin{equation}
\delta \left( \sqrt{u^{\nu }u_{\nu }}-1\right) \mathrm{d}^{4}\mathbf{%
u\rightarrow
}\frac{\mathrm{d}u_{1}\mathrm{d}u_{2}\mathrm{d}u_{3}}{u^{0}}.
\end{equation}%
However, in the gyrokinetic transformation the four-velocity is
expressed via Eqs. (\ref{um})-(\ref{uo}) as
\begin{equation}
u_{\mu }=w\left( l_{\mu }^{\prime }\cos \phi +l_{\mu }^{\prime
\prime }\sin \phi \right) +{u_{\parallel }}l_{\mu }+u_{o}\tau
_{\mu },
\end{equation}%
so that $\mathrm{d}^{4}\mathbf{u=}w\mathrm{d}w\mathrm{d}\phi \mathrm{d}%
u_{\parallel }\mathrm{d}u_{o}$ [the sign is positive due to Eq.(\ref{orderb}%
)], while the partial integration over $\mathrm{d}u_{o}$ leads to
\begin{equation}
\delta \left( \sqrt{u^{\nu }u_{\nu }}-1\right) \mathrm{d}^{4}\mathbf{%
u\rightarrow }\frac{w\mathrm{d}w\mathrm{d}\phi \mathrm{d}u_{\parallel }}{%
u_{o}},
\end{equation}%
where $u_{o}=\sqrt{1+w^{2}+{u_{\parallel }}^{2}}.$ As a result,
the expression for components of the current density can be
rewritten as
\begin{equation}
j^{\mu }=c\sum_{\alpha }q_{\alpha }\int \left( w\left( l^{\prime
\mu }\cos \phi +l^{\prime \prime \mu }\sin \phi \right)
+{u_{\parallel }}l^{\mu }+u_{o}\tau ^{\mu }\right) f_{\alpha
}\left( \mathbf{x},\mathbf{u}\right)
\frac{w\mathrm{d}w\mathrm{d}\phi \mathrm{d}u_{\parallel
}}{\sqrt{-g}u_{o}}. \label{jm}
\end{equation}%
Further, the distribution function $f_{\alpha }$ is expressed as
the function of the gyrokinetic variables
\begin{equation}
f_{\alpha }=f_{\alpha }\left( x^{\prime \mu },\widehat{\mu
},u_{\parallel }\right) ,
\end{equation}%
and it is necessary to transform it back to particle coordinates
before integrating, as in Eq.(\ref{jm}) the particle position
$\mathbf{x,}$\ rather than its gyrocenter position
$\mathbf{x}^{\prime },$\ is kept constant while integrating over
the particle velocity. This makes it convenient to rewrite
Equation (\ref{sp}) as
\begin{equation}
\frac{\partial }{\partial x^{\nu }}\left( \sqrt{-g}F^{\mu \nu }\right) =-%
\frac{4\pi }{c}j^{\mu }\sqrt{-g}=Q^{\mu }(\mathbf{x}),
\end{equation}%
where the right-hand side is also evaluated at $\mathbf{x}.$ Then
\begin{equation}
Q^{\mu }(\mathbf{x})=-4\pi \sum_{\alpha }q_{\alpha }\int \left[
w\left(
l^{\prime \mu }\cos \phi +l^{\prime \prime \mu }\sin \phi \right) +{%
u_{\parallel }}l^{\mu }+u_{o}\tau ^{\mu }\right] f_{\alpha }\left( \mathbf{x}%
-\sum_{i=1}\varepsilon ^{i}\mathbf{r}_{i}\right) \frac{w\mathrm{d}w\mathrm{d%
}\phi \mathrm{d}u_{\parallel }}{u_{o}}.
\end{equation}.%

\section{Conclusion}

A closed set of relativistic gyrokinetic equations, consisting of
the collisionless gyrokinetic equation and the averaged Maxwell's
equations, is derived for an arbitrary four-dimensional coordinate
system. \\
In several respects the theory here developed represents a
significant improvement with respect to kinetic equations derived
by other authors. The present covariant kinetic theory adopts a
set of hybrid gyrokinetic variables, two of which include the
Lorentz-invariant magnetic moment and gyrophase angle.  The
theory, allows $E/B\sim O(1)$ and therefore permits relativistic
drifts ($V_d\sim c$) an moreover takes into account nonlinear
effects of the EM wave-fields. Moreover, since the gyrokinetic
transformation is obtained to the second order in terms of the
ratio of the Larmor radius to the inhomogeneity scale, the theory
can be applied also to the investigation of finite-Larmor radius
effects. Another interesting aspect is that in the present theory
the wave field is no longer limited in frequency and the
wavelength, i.e., $\omega /\Omega _{c}\sim O(1),$ $k_{\Vert }\rho
_{L}\sim O(1)$, so that the class of admissible waves is broader
than the usual ``drift-Alfven perturbations'' and can include the
magneto-sonic waves, for example.


\begin{theacknowledgments}
Work developed in the framework of the PRIN Research Program
``Programma Cofin 2002: Metodi matematici delle teorie
cinetiche''( MIUR Italian Ministry) and conducted via the
cooperation program between the Trieste University, Italy, and the
Budker Institute of Nuclear Physics, Novosibirsk, Russia. The
research has been partially supported (for A.B. and P.N.) by the
National Group of Mathematical Physics of INdAM (Istituto
Nazionale di Alta Matematica), (P.N) by the INFN (Istituto
Nazionale di Fisica Nucleare), Trieste (Italy) and (M.T.) by the
Consortium for Magnetofluid Dynamics, University of Trieste, Italy
and (A.B.) by the University of Trieste.
\end{theacknowledgments}



\bibliographystyle{aipproc}
\bibliography{sample}

\begin{thebibliography}{8}
\bibitem{Mohanti}  J. N. Mohanty and K. C. Baral, Phys. Plasmas {\bf 3}, 804 (1996) and references therein.

\bibitem{Beklem} A.Beklemishev and M.Tessarotto, Phys. Plasmas {\bf 6}, 4487 (1999).

\bibitem{Beklem2004}  A.Beklemishev and M.Tessarotto, submitted (2004).

\bibitem{Pozzo1998} Pozzo M., Tessarotto M., Phys.
Plasmas, \textbf{5,} 2232 (1998).

\bibitem{Morozov 1966} A.I. Morozov and L.S. Solov'ev, in {\it Reviews of
Plasma Physics, }Edited by Acad. M.A. Leontovich (Consultants
Bureau, New York, 1966), Vol. 2, p. 201.

\bibitem{Gardner1959} C.S. Gardner, Phys. Rev. {\bf 115}, 791 (1959).

\bibitem{Weitzner 1995} H. Weitzner, Phys. Plasmas, {\bf 2}, 3595 (1995).


\bibitem{Littlejohn1979} R.G. Littlejohn, \ J. Math. Phys. {\bf 20}, 2445
(1979).

\bibitem{Littlejohn1981} R.G. Littlejohn, Phys.Fluids{\bf \ 24}, 1730 (1981).

\bibitem{Littlejohn1983} R.G. Littlejohn, \ J. Plasma Phys.{\bf \ 29}, 111
(1983).

\bibitem{Grebogi} C. Grebogi, R. G. Littlejohn, Phys. Fluids {\bf 27}, 1996
(1984);\

\bibitem{Littlejohn1985} R. G. Littlejohn, Phys. Fluids {\bf 28}, 2015
(1985).

\bibitem{Boozer1996} A. H. Boozer, Phys. Plasmas {\bf 3}, 3297
(1996).

\bibitem{Cooper1997} W. A. Cooper, Plasma Phys. Control. Fusion {\bf 39}, 931
(1997).

\bibitem{Brizard1999} A.J.Brizard, Phys.Plasmas {\bf 6}, 4548 (1999).

\bibitem[Littlejohn (1984)]{LJ} Littlejohn R. G., Phys. Fluids \textbf{%
27}, 976 (1984).

\bibitem[Boozer (1996)]{BO} Boozer A. H., Phys. Plasmas, \textbf{3,}
3297 (1996).

\bibitem[Cooper (1997)]{CO} Cooper W. A., Plasma Phys. Control. Fusion,
\textbf{39,} 931 (1997).

\bibitem[Brizard \& Chan (1999)]{Br} Brizard A.J., Chan A.A., Phys.
Plasmas, \textbf{6}, 4548 (1999).

\bibitem[Landau \& Lifshits(1975)]{LL1} Landau L. D., Lifshits E. M. 1975,
The Classical Theory of Fields, 4th ed., Pergamon, Oxford, 1975.
\end{thebibliography}

\end{document}